\documentclass[twocolumn]{aastex701}

\usepackage[dvipsnames]{xcolor}
\usepackage{float}  
\usepackage{braket}
\usepackage{amsmath} %

\begin{document}

\title{Eccentricity Distribution
for Dynamical Gravitational-Wave Merger Channels in The Pinhole Regime
}

\author[orcid=0000-0002-2728-0132,sname='Rozner']{Mor Rozner} 
\affiliation{Institute for Advanced Study, Einstein Drive, Princeton, NJ 08540, USA}
\affiliation{Institute of Astronomy, University of Cambridge, Madingley Road, Cambridge CB3 0HA, UK}
\affiliation{Gonville \& Caius College, Trinity Street, Cambridge CB2 1TA, UK}
\email[show]{morozner@ias.edu}

\author[orcid=0000-0002-6714-5429,sname='Clarke']{Teagan A.~Clarke} 
\affiliation{Department of Physics, Princeton University, Princeton, New Jersey, 08544, USA}
\email[]{teagan.clarke@princeton.edu}

\author[orcid=0000-0002-4181-8090,sname='Romero-Shaw']{Isobel M. Romero-Shaw} 
\affiliation{Gravity Exploration Institute, School of Physics and Astronomy, Cardiff University, Cardiff, CF24 3AA, UK}
\email[]{isobel.romeroshaw@gmail.com}

\author[orcid=0000-0003-0607-8741,sname='Romero-Shaw']{Johan Samsing} 
\affiliation{Niels Bohr International Academy, The Niels Bohr Institute, Blegdamsvej 17, DK-2100, Copenhagen, Denmark
}
\email[]{jsamsing@gmail.com}

\begin{abstract}
We argue that various dynamical astrophysical black hole merger channels are expected to result in a common eccentricity distribution at gravitational-wave (GW) frequencies relevant for
LIGO/Virgo/KAGRA (LVK) in the high eccentricity limit. This follows from the large separation
of scales between the GW regime required for creating eccentric mergers in LVK, and the underlying astrophysical formation environment. Our analytical solution shows exceptional agreement
with numerical studies. This finding has important implications for both theoretical studies and
ongoing searches for eccentric GW sources.\end{abstract}



\section{Introduction}

The first detection of gravitational waves (GWs) marked the beginning of a new era in astrophysics \citep{Abbott2016, GW150914_GWB:2016}. 
A decade later, more than $200$ BBH (binary black hole) mergers have been confirmed by the LIGO-Virgo-KAGRA (LVK)
collaboration \citep{adv_ligo_2015, AdvancedVirgo, 2020_Kagra} with a variety of
properties \citep[e.g.,][]{GWTC1_2019, GWTC2, GWTC-2_RnP, GWTC2.1_2024, gwtc3, O3b_population, gwtc4, 2025arXiv250818083T},
which not only have enabled statistical studies, but also revealed physical systems that were once thought to be peculiarities.
Many GW observables of merging black holes (BHs), such as mass and spin distributions, are somewhat degenerate across the proposed formation
channels, which makes them hard to disentangle.
However, a few recent outliers might have provided some hints, including {GW200208\_222617} and GW200105 that show signs
of eccentricity \citep{Romero-Shaw2022, Gupte2025, 2025PhRvD.112f3052R, Morras2025, Planas2025, Jan2026},
GW190521 and GW231123 that are characterised by high masses
well above the theoretically expected pair instability mass cut-off \citep{Abbott2020_190521, 2025ApJ...993L..25A},
and GW241110 that shows a mass ratio $\sim{1:2}$ with the second spinning around $\sim 0.7$, indicating that the most massive BH itself
could be a product of a previous BBH merger \citep{2025ApJ...993L..21A}. Such hierarchical mergers are especially useful for
providing constraints on various cluster channels \citep[e.g.,][]{2021ApJ...923..126S, Kimball2021, Antonini2025, Farah2026, Vijaykumar2026}.

Eccentric mergers are a smoking gun for dynamical formation mechanisms in general, including
single-single encounters \citep[e.g.,][]{OLeary2009, 2020PhRvD.101l3010S,GondanKocsis2021}, binary-single
encounters \citep[e.g.,][]{2017ApJ...840L..14S, 2018ApJ...863....7R, Samsing18,Samsing22,GinatPerets2021,
Ginat2023}, binary-binary encounters \citep[e.g.,][]{2019ApJ...871...91Z}, hierarchical interactions \citep[e.g.,][]{Wen:2003bu,
AntoniniPerets2012, 2016ApJ...831..187A, 2016ARA&A..54..441N,2017arXiv170609896H, Grishin2018,
Randall2018,
Su2025,Dorozsmai2026}, gas-assisted mergers \citep[e.g.,][]{2020ApJ...898...25T,RoznerPerets2022,Rowan2025}, and close encounters in
discs \citep[e.g.,][]{Li2022, 2022Natur.603..237S, Rom2024, Fabj24}.
Their distribution and occurrence rates are expected to provide one of the best and well-defined ways of distinguishing
between dynamical and isolated binary formation channels both on a population
level \citep[e.g.,][]{2014ApJ...784...71S, 2017ApJ...840L..14S, Samsing18a, Samsing2018, Samsing18, 2018ApJ...855..124S, 2018MNRAS.tmp.2223S, 2018PhRvD..98l3005R, 2019ApJ...881...41L,2019ApJ...871...91Z, 2021ApJ...921L..43Z,
Romero-Shaw2022, 2022ApJ...940..171R,Stegmann2026}, as well as through individual
sources \citep[e.g.][]{2025ApJ...990..211S, 2025CQGra..42u5006T}. This promise has sparked a major effort to create
accurate eccentric waveform models \citep[e.g.,][]{TEOBResumS, Gamboa2025, Planas2026}, GW template
banks \citep{Wang2025, Phukon2025}, and optimised search techniques for such eccentric GW sources  \citep{Abac2024_eccentric, Phukon2025b}.
For these searches, which are still in a very initial stage,
it is highly useful to have a handle on the expected eccentricity distribution. Currently, observational analyses typical use
a uniform or a log-uniform eccentricity
distribution, $P(e)\propto 1/e$ \citep[e.g.,][]{RomeroShaw2020, Romero-Shaw2022, Gupte2025, 2025PhRvD.112f3052R, Morras2025, Planas2025, Jan2025};
however, we show here that this is not a natural prior.

In this \textit{Letter}, we illustrate that in the \textit{high eccentricity regime} at
frequencies relevant for LVK ($f\geq 10 \ \rm{Hz}$), various dynamical merger channels
are expected to follow the same eccentricity distribution. At the post-Newtonian (PN) level, this distribution
takes a closed analytical form that is unique and clearly different from typically assumed distributions, including $P(e) \propto 1/e$.
Our findings are based on the argument that the astrophysical scales associated with BBHs at their assembly are much bigger
than the ones required for eccentric mergers, which essentially makes the formation of
eccentric mergers a stochastic process rather than fine-tuned with memory
from the underlying astrophysical scales. We denote this as the
{\it `pinhole regime'} \footnote{The term 'pinhole' is commonly used in the context of dynamics around supermassive black holes, e.g. \citep{LightmanShapiro1977}, which inspired our notation here}, as further illustrated in Fig.~\ref{fig:Ill_pinhole}.
We compare with a few examples from fully dynamical N-body simulations, including chaotic binary-single
scatterings and hierarchical triples, and indeed, find that our analytical solution correctly predicts the
eccentricity distribution. Our suggested distribution
should serve as the best benchmark for
eccentricity searches and analyses in the high eccentricity regime that currently defines the resolvable limit by LVK.

\section{Assembly of Black Hole Binaries}\label{sec:Assembly of Black Hole Binaries}

The assembly of a high-eccentricity BBH merger generally results from two BHs that encounter
each other at a pericentre distance, $r$, after which the evolution becomes strongly GW-driven until merger.
During the inspiral, the BBH will gradually circularise, but can still, in some cases, maintain residual
eccentricity when it enters the detectable GW frequency regime, typically $\geq 10 \ \rm{Hz}$ \citep{Harry_2022} for LVK.
In this \textit{Letter}, we operate with two different frequencies: the
peak frequency, $f_p= \pi^{-1} \sqrt{GM/r_f^3}$, where $M$ is the total mass of the BBH,
$r_f$ is the pericentre distance; and the orbital (Keplerian) frequency, $f_T=  \pi^{-1} \sqrt{GM/a_f^3}$,
where $a_f$ is the BBH semi-major axis (SMA).

The question that we are aiming to solve is: what is the orbital eccentricity distribution
at a given frequency $f$, and how does it depend on the underlying formation mechanism? 
As the orbital evolution of highly eccentric BBHs is essentially dictated by the change of
angular momentum and energy over the first pericentre passage, the relevant initial conditions to start with are the distribution of pericentre distances $r$.
The calculation then boils down to deriving the distribution of $r$ and mapping it to the distribution
of eccentricity at a given frequency $f$. We will proceed with this calculation in the following sections.

\subsection{Initial Conditions}\label{subsec:initial_conditions}

\begin{figure}
    \centering
    \includegraphics[width=1\linewidth]{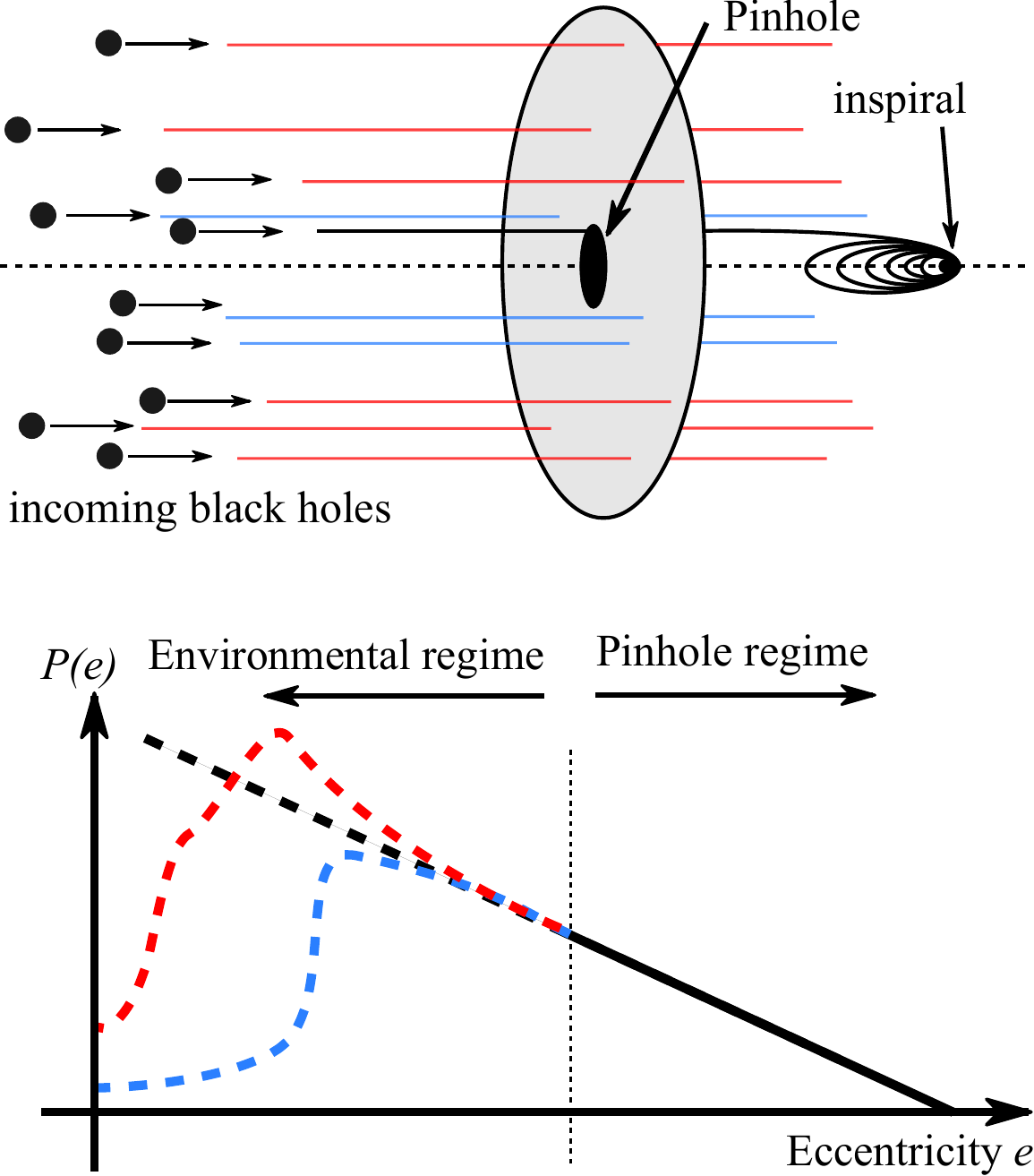}
    \caption{{\bf The Pinhole Regime.}
    {\it Top:} For a given encounter to result in an eccentric
    merger, the two interacting BHs must pass each other at a relatively small pericentre distance.
    This restriction can be mapped to a corresponding small area
    at the encounter surface, which we refer to as the `pinhole'.
    As the size of this area is generally much smaller than any of the underlying astrophysical scales, the distribution of BHs
    passing through this area is approximately random, which maps to our proposed eccentricity distribution,
    as described in Sec. \ref{sec:universal}.
    {\it Bottom:} At high eccentricity, all of the channels are expected to converge towards the same functional form, which is a result of the separation of scales. At lower eccentricity, the astrophysical environment will start to show up, however,
    LVK is not expected to resolve this limit yet with the current sensitivity.}
    \label{fig:Ill_pinhole}
\end{figure}

The scales associated with most dynamical astrophysical eccentric merger channels are much larger than the typical pericentre distance required to
form a BBH with measurable eccentricity in LVK \citep[e.g.,][]{Samsing18}.
This separation of scales leads us to introduce the \textit{`pinhole regime'}, which describes our setup where the
incoming BH has to go through a `hole' with radius $b$, i.e. the pinhole, to create the eccentric BBH merger, as illustrated in Fig. \ref{fig:Ill_pinhole}.
Because of this separation of scales, the spatial distribution of the BHs at the pinhole will, to leading order, be stochastic, 
i.e. randomly distributed across the pinhole surface.
To see how this propagates to observables, we start by relating the impact parameter of
the incoming BH, $b$, to the corresponding pericentre distance, $r$, using Kepler's equation, 
\begin{align}
b = r\sqrt{1+{2GM}/{(rv_{\infty}^2)}} \propto \sqrt{r}, 
\end{align}
where $M$ is the mass of the binary, and $v_\infty$ is the velocity of the incoming BH at infinity,
assuming that the velocity at pericentre is $\gg v_{\infty}$, which holds
in the close encounter regime.
With this relation, we can relate the distribution of incoming BHs, $P(b)$,
to the corresponding distribution of pericentre distances, $P(r)$, as
\begin{align}\label{eq:P(r)}
P(r) = P(b) \left|{db}/{dr} \right|,
\end{align}
where we throughout the {\it Letter} refer to the distribution of variable $\mathcal{A}$ by $P(\mathcal{A})$.
The distribution for $P(b)$ follows from the random encounter regime that describes
the pinhole regime, which implies that the number of BHs entering within a radius $b$ is
simply proportional to the enclosed area, $\propto b^2$ \footnote{We note that for substantially different geometries,
such as close to two-dimensional disc-like systems \citep[e.g. AGN discs][]{Samsing22}, $P(b)$ will change \citep{2025PhRvD.112f3052R},
and the derivation should be repeated with the relevant distribution.}. From this follows that
$P(b) \propto b$, and $|db/dr| \propto r^{-1/2}$.
Substituting these terms into Eq. \eqref{eq:P(r)} above, we find that
\begin{align}
P(r) \propto \rm{constant}.
\label{eq:Pr_const}
\end{align}
This distribution is generally valid for processes where the single incoming BH is isotropically distributed at infinity, such as in most
spherical-symmetric dense environments, as further discussed
in Sec. \ref{sec:astro_examples}.

\section{Eccentricity Distribution}\label{sec:universal}

\begin{figure}
    \centering
    \includegraphics[width=1.\linewidth]{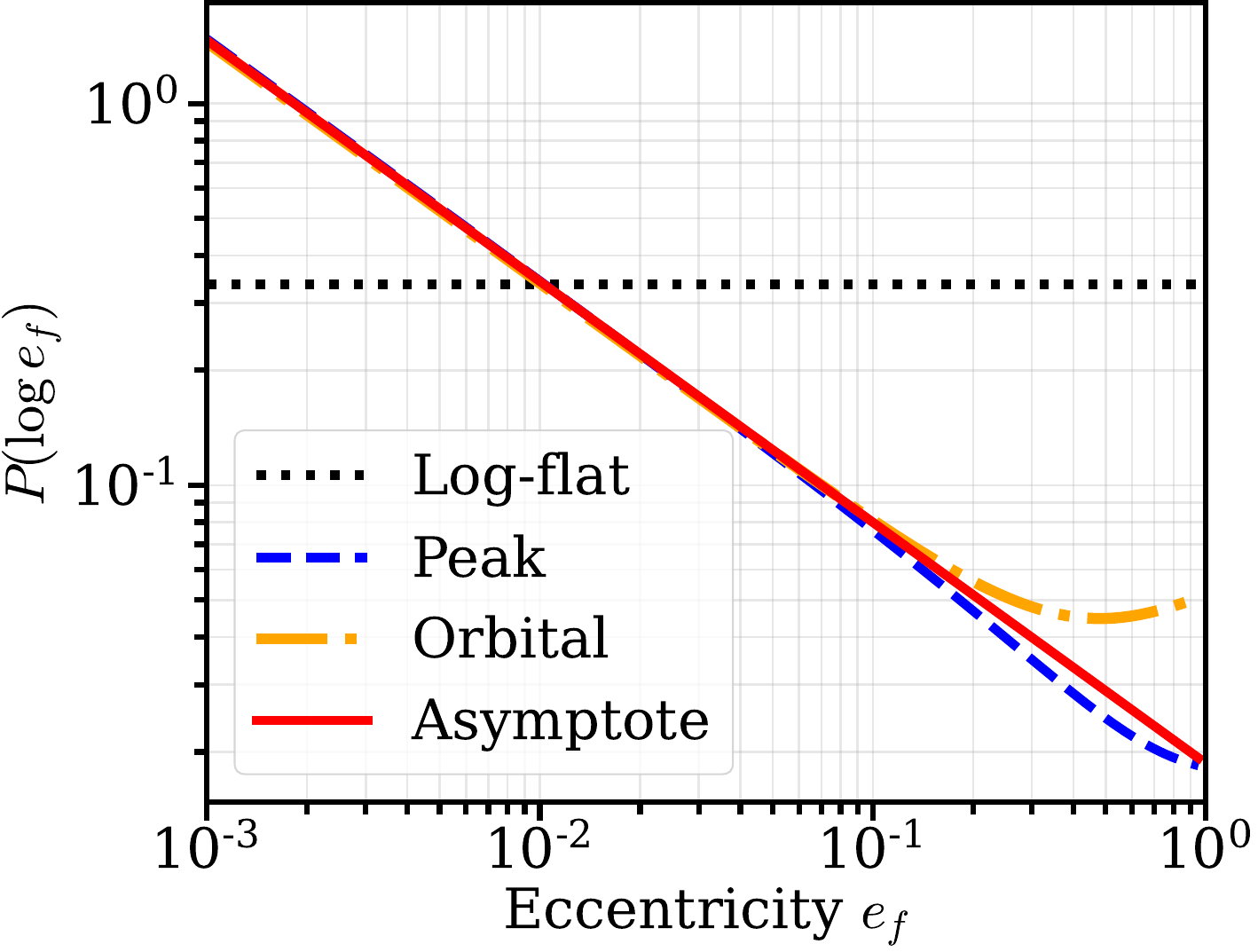}
    \caption{{\bf Analytical Eccentricity Distributions.} The probability density function of the log of eccentricity, $P(\log e) \propto eP(e)$, as calculated in Eqs. \eqref{eq:P_peak} and \eqref{eq:PT} for the peak and orbital frequencies correspondingly (in dashed and dashed-dotted lines), compared to the asymptotic solution (solid line) and log-flat distribution (dotted line). The distributions will break below a characteristic eccentricity, $e_c$, set by the astrophysical environment, and at high eccentricity when finite effects are included (see Sec. \ref{sec:universal}).
    It should be noted that the distributions at peak and orbital frequencies relate to different distributions in terms of eccentricity, which is defined accordingly.
    }
    \label{fig:placeholder}
\end{figure}

The distribution of orbital eccentricity, $e_f$, evaluated at
GW frequency $f$, is related to the distribution in initial pericentre distance, as follows,
\begin{align}
P(e_f) = P(r) \left|{dr}/{de_f}\right| \propto \left|{dr}/{de_f}\right|
\label{eq:Pef_drde}
\end{align}
where the last step follows from Eq. \eqref{eq:Pr_const}.
The relation between $r$ and $e_f$ can be found through the
relation, $a(e) \propto e^{12/19}(1+121e^2/304)^{870/2299}/(1-e^2)$, between the binary SMA, $a$, and
the eccentricity, $e$, given by \citep{Peters:1964bc}. From this, one can now write the initial pericentre
distance, $r$, as a function of the BBH eccentricity $e_f$, at GW frequency $f$ \cite[e.g.,][]{2025PhRvD.112f3052R}. When considering peak frequency, 
\begin{equation}
r \approx C\frac{1+e_f}{e_f^{12/19}} \left(1 + \frac{121}{304}e_f^2 \right)^{-\frac{870}{2299}} M^{1/3} f_p^{-\frac{2}{3}},
\label{eq:r_EM}
\end{equation}
and when considering orbital frequency,
\begin{equation}
r \approx C\frac{1-e_f^2}{e_f^{12/19}} \left(1 + \frac{121}{304}e_f^2 \right)^{-\frac{870}{2299}} M^{1/3} f_T^{-\frac{2}{3}},
\label{eq:r_EM2}
\end{equation}
where $C = \frac{1}{2}(G/\pi^2)^{1/3}$, $f_p$ refers to the GW peak frequency, and $f_{T}$ refers to the orbital frequency,
as defined in Sec. \ref{sec:Assembly of Black Hole Binaries}. Both
of these definitions are currently being used in the literature, among other similar definitions \citep[e.g.,][]{Wen:2003bu,2018PhRvL.120s1103K,
2019PhRvD..99f3006S, Hamers2021}. 
With these relations, we can now write out the eccentricity distribution evaluated at GW frequency
$f$, using Eqs. \eqref{eq:Pef_drde}, \eqref{eq:r_EM} and \eqref{eq:r_EM2}, from which follows for the distribution using peak frequency
\begin{align}\label{eq:P_peak}
P(e_{f,p}) \propto \frac{192-112e_{f,p}+168e_{f,p}^2+47e_{f,p}^3}{e_{f,p}^{31/19}(304+121e_{f,p}^2)^{3169/2299}},
\end{align}
and when using orbital frequency,
\begin{align}\label{eq:PT}
P(e_{f,T}) \propto \frac{96+292e_{f,T}^2 + 37e_{f,T}^4}{e_{f,T}^{31/19}(304+121e_{f,T}^2)^{3169/2299}},
\end{align}
where $P(e_{f,p})$ and $P(e_{f,T})$ denote the distribution evaluated at the GW peak frequency ($f_p$) and the orbital frequency ($f_T$), respectively.
As can be seen, common to the two distributions is that they both asymptote to
\begin{align}
P(e_f) \propto e_f^{-31/19}.
\label{eq:Pefe3119}
\end{align}
Or equivalently, $P(\log e_f)\propto e_f^{-12/19}$, where we have used $P(\log e) \propto eP(e)$.
In addition, we see that they do not depend on the actual value for the frequency $f$ or the BH
masses, as long as our assumptions hold and finite size effects are not included.
Fig. \ref{fig:placeholder} shows the analytical solutions from Eq. \eqref{eq:P_peak} and \eqref{eq:PT} together with the
often assumed prior, $P(e) \propto 1/e$ (log-flat).
While $P_p$ gradually decreases with $e_f$, one notes that $P_T$ surprisingly has a significant upturn
towards higher eccentricities. This behaviour is correct in the orbit-averaged point-mass
adiabatic $2.5 \rm{PN}$ framework we consider, but requires
numerical-relativity (NR) simulations to correctly describe. This can, e.g., be seen by considering
Eq. \eqref{eq:r_EM2} (see also Appendix \ref{app:B}),
where $r/(2GM/c^2) \lesssim 10$ for $e_f \gtrsim 0.2$ at $M=40 \ M_{\odot}, f_T=20\ \rm{Hz}$, which
illustrates that even modest eccentricities map to highly relativistic encounters.
Below, we put our results in relation to astrophysical cases.

\begin{figure}
    \centering
    \includegraphics[width=1\linewidth]{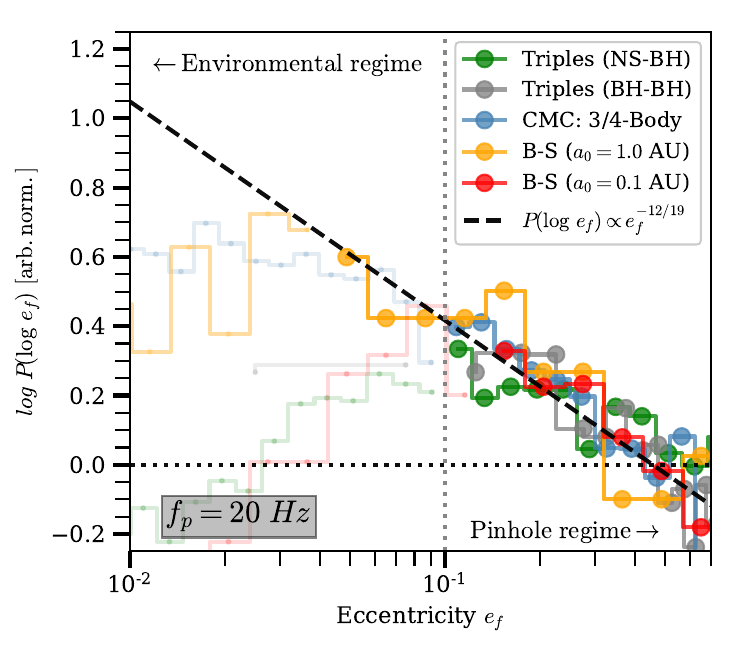}
    \includegraphics[width=1\linewidth]{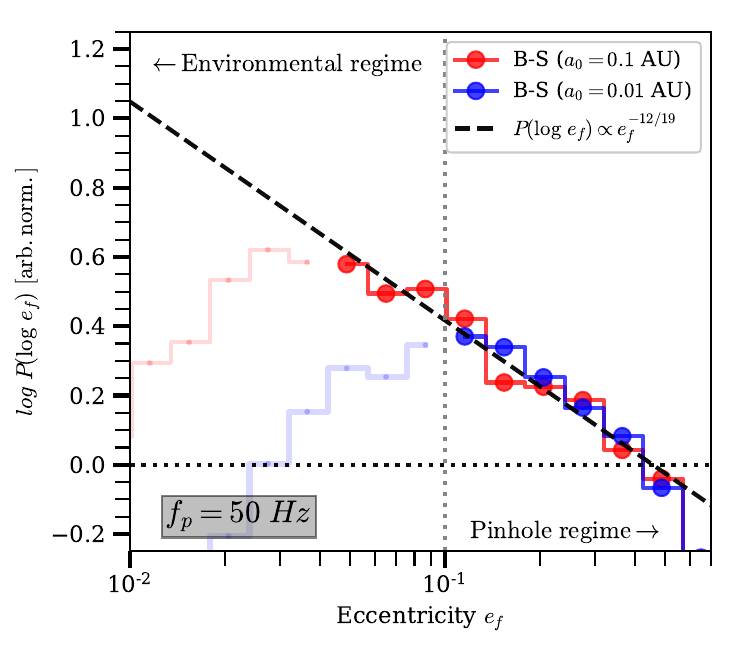}
    \caption{{\bf Astrophysical Eccentricity Distributions.}
    Eccentricity distributions for different assembly mechanisms taking place in representative astrophysical settings.
    {\it Top panel:} Distributions evaluated at $f_p = 20 \  \rm{Hz}$. In {\it green} is shown results for ZLK induced neutron star-BH mergers
    from \cite{Stegmann2025} (Triples (NS-BH)), where in {\it grey} is shown results for ZLK induced BH-BH mergers from \cite{2014ApJ...781...45A}
    (Triples (BH-BH)). 
    In {\it light blue} we show results from Cluster Monte Carlo (CMC) simulations \cite{2021ApJ...921L..43Z}.
    In {\it orange} and in {\it red} are shown results from equal mass $m = 20 \ M_{\odot}$ binary-single BH interactions
    with initial SMA (semimajor axis) values $a_0 = 1 \ \rm{AU}$ and $a_0 = 0.1\ \rm{AU}$, respectively (B-S). Data from the binary-single interactions
    have been shaded below their characteristic eccentricity, $e_c$, given by Eq. \eqref{eq:ec_BS}, where
    the 
    triples datasets have been shaded below $e = 0.1$.
    {\it Bottom panel:} Distributions at $f_p = 50  \ \rm{Hz}$ for binary-single interactions with
    initial SMA values $a_0 = 0.1\ \rm{AU}$ and $a_0 = 0.01\ \rm{AU}$. The better statistics
    in this limit clearly show how the BBHs follow our theoretical prediction.
    }
    \label{fig:p(e)_data}
\end{figure}

\section{Astrophysical Scenarios}\label{sec:astro_examples}

We compare our analytical predictions with distributions from distinct astrophysical formation channels of eccentric mergers, with results summarised in Fig. \ref{fig:p(e)_data}.

\subsection{The Pinhole Limit}

The first channel is the assembly of
BBHs in dense stellar clusters, where high eccentricity is reached through strong and chaotic
binary-binary and binary-single interactions \citep[e.g.,][]{Samsing14, 2016PhRvD..93h4029R, Samsing18, 2018ApJ...855..124S, 2018PhRvL.120o1101R, 2019ApJ...871...91Z}.
In this case, the general finding is that the constant perturbation of the BBH angular momentum leads to
a distribution in orbital eccentricity, $P(e) = 2e$, also referred to as a thermal distribution \citep{Jeans1919,Ambartsumian1937}.
Here the distribution of pericentre distance can be calculated
as $P(r) = P(e)|de/dr| \approx 2/a \propto \rm{constant}$, where we have assumed that the pericentre distance leading to a high eccentricity merger is much smaller than the initial BBH SMA. As $P(r) \propto \rm{constant}$, this
channel is therefore expected to follow our pinhole distribution.
To explore this, we show in Fig. \ref{fig:p(e)_data} results from controlled 
scatterings between a BBH with SMA $a_0$ and an incoming single BH,
for $a_0 = 1\ \rm{AU},\ 0.1\ \rm{AU},\ 0.01\ \rm{AU}$, assuming equal masses of $m = 20 \ M_{\odot}$ \citep[e.g.][]{Samsing14}.
Each of these configurations is simulated $250,000$ times, assuming an isotropic scattering environment. We further show data from Cluster Monte Carlo (CMC)
simulations \cite{2021ApJ...921L..43Z} (Fig. 1), that include both binary-single and binary-binary interactions for a large ensemble of realistic globular clusters. As seen in the figure, there is excellent agreement in the high eccentricity limit, i.e. the pinhole limit, between our asymptotic scaling, $P(e) \propto e^{-31/19}$,
and numerical data. The $f_p = 50\ \rm{Hz}$ case (bottom plot) is included as this leads to an increase in statistics that allows us to clearly illustrate that pinhole solution is indeed achieved.

The second case is BBHs merging in hierarchical triples undergoing von Zeipel-Lidov-Kozai (ZLK) oscillations  \citep[e.g.][]{VonZeipel1910, 1962P&SS....9..719L,1962AJ.....67..591K,Blaes2002, 2013ApJ...773..187N, 2014ApJ...785..116L, 2016ApJ...816...65A, 2016MNRAS.456.4219A, 2017ApJ...836...39S, 2018ApJ...864..134R, 2019ApJ...883...23H,
2020ApJ...903...67M, 2021MNRAS.502.2049L, 2022MNRAS.511.1362T},
where the high eccentricity tail has been argued to mainly originate from non-secular configurations, and decoupled from secular evolution \citep[e.g.,][]{2014ApJ...781...45A}.
These can form through e.g. binary-binary interactions \cite[e.g.][]{2019ApJ...871...91Z} or through
later fly-by perturbations \citep[e.g.][]{2020MNRAS.498.4924M}.
In this non-secular regime, the non-adiabatic evolution is expected to lead to a near uniform distribution in $J^2$, where $J$ is the angular momentum, which again leads to $P(r) \approx 2/a \propto \rm{constant}$, as also argued in \cite[e.g.,][]{2012arXiv1211.4584K}. To test this, we use for the comparison results from \cite{2014ApJ...781...45A} (N-body results, Fig.~5),
and \cite{Stegmann2025} (N-body results, Fig.~7), where using the N-body results is crucial, as the correct stochastic sampling at high eccentricity cannot be achieved using the standard secular evolution equations, due to the decoupling from secular evolution \citep[e.g.,][]{2012arXiv1211.4584K, 2014ApJ...781...45A}. We see great agreement in Fig. \ref{fig:p(e)_data} with our asymptotic solution.
We caveat that some mergers could originate from bursting sources, in which a merger takes place on very short timescales, typically shorter
than pericentre passage time \citep[e.g.][]{Xuan2024,Xuan2026}, or from the empty loss-cone (which will require taking into account a
different $P(r)$ before carrying on with the rest of the derivation). These cases require a dedicated, careful study related to
near-plunge-orbit dynamics, and are out of the scope of the current letter. As long as we are in the full loss-cone, i.e., pinhole, regime,
eccentric mergers produced by this channel should obey the pinhole eccentricity distribution derived above.     

We finally note that the often considered single-single GW capture
channel \citep[e.g.][]{OLeary2009, 2020PhRvD.101l3010S} will, by definition, give rise to the distributions we derive, as such systems generally originate from isotropically encounters. We therefore leave out simulations of this process.

\subsection{The Environmental limit}

While we find great agreement in the high eccentricity limit between our considered channels and our formalism,
it is clear that the distributions start to deviate from the power law solution below a
certain critical eccentricity, $e_c$. This eccentricity depends on the environment \citep[e.g.][]{2026arXiv260329019G}.
In a typical astrophysical setting, this $e_c$ corresponds roughly to the pericentre distance at which the change
in orbital parameters by the emission of GWs is comparable to their initial values.
For example, the maximum pericenter distance a BBH can form through a single-single encounter between
two initially unbound BHs, is where the energy
radiated at pericenter, $E_{GW} \propto m^{9/2}r^{-7/2}$, is similar to the initial energy at
infinity, $E_{0} \propto mv_{env}^{2}$,
\begin{equation}
r_c \approx R_{s}(c/v_{\rm{env}})^{4/7},
\label{eq:rcap_env}
\end{equation}
where $R_s = 2GM/c^2$, $c$ is the speed of light, and $v_{env}$ is the characteristic velocity dispersion
of the environment \citep[e.g.][]{Hansen:1972il, 1989ApJ...343..725Q, OLeary2009, Samsing14, Samsing18, 2020PhRvD.101l3010S}.
By substituting this expression for $r$ into Eq. \eqref{eq:r_EM} or \eqref{eq:r_EM2}, one can now extract the corresponding
eccentricity,
\begin{equation}
    e_c  \approx 0.01\times \left(\frac{M}{50 \ M_{\odot}}\right)^{-\frac{19}{18}} \left(\frac{v_{\rm{env}}}{50\ \rm{km} \ \rm{s}^{-1}}\right)^{\frac{19}{21}} \left(\frac{f}{10\ \rm{Hz}}\right)^{-\frac{19}{18}}
    \label{eq:ec_venv}
\end{equation}
where we have assumed that $e_c \ll 1$. For initially bound BBHs, such as the one driven to merger through triple interactions, the
characteristic pericenter distance is instead given by when the radiated energy, $E_{\rm{GW}}$, is similar to the BBH
orbital energy $E_{\rm{orb}} \propto m^2/a$,
\begin{equation}
r_c \approx R_{s}(a/R_{s})^{2/7},
\label{eq:rcap_BS}
\end{equation}
where $a$ is the initial SMA of the BBH \citep[e.g.][]{Samsing18}.
Similarly, one can map this to a corresponding eccentricity in the limit $e_c \ll 1$,
\begin{equation}
    e_c  \approx 0.1 \times \left(\frac{M}{40M_{\odot}}\right)^{-\frac{38}{63}} \left(\frac{a}{1\ \rm{AU}}\right)^{-\frac{19}{42}} \left(\frac{f}{10\ Hz}\right)^{-\frac{19}{18}}.
    \label{eq:ec_BS}
\end{equation}
In Fig. \ref{fig:p(e)_data} we have shaded the data from the binary-single interactions with $e_f < e_c$ estimated by Eq. \eqref{eq:ec_BS}.
As seen, our estimator provides an excellent division between
the environmental regime ($e_f \lesssim e_c$) and the pinhole regime ($e_f \gtrsim e_c$). A similar critical eccentricity can be derived for ZLK-induced mergers \citep[e.g.][]{Stegmann2025}. Further discussions are given in
Sec. \ref{sec:Observational Implications}.

\section{Conclusions}\label{sec:conclusions}

We suggest in this {\it Letter} that various dynamical formation channels that produce eccentric mergers,
from chaotic few-body systems to hierarchical triples,
are expected to result in a common eccentricity distribution in the high eccentricity limit at frequencies relevant for LVK observations.
We derive this distribution using three independent methods: analytically, semi-analytically (continuity equation), and by comparing our results to numerical simulations of three dynamical channels, all of which show great agreement.
We argue that this stems from a large separation of scales between very close binaries in the final stages before an eccentric merger and the much larger scales at their formation. 
In the final stages of the merger, the evolution of the binaries is dictated by Peters equation \citep{Peters:1964bc} solely, retaining no memory 
of their birth environment or the formation process, which characterises the
final stages of a merger. We term this the `pinhole regime', and show analytically that the common distribution
should follow $P(e) \propto e^{-31/19}$.

We find excellent agreement with results from simulations describing chaotic interactions
in dense globular clusters and ZLK interacting triples, as further discussed and shown in
Fig. \ref{fig:p(e)_data}. This finding can be used as a prior for LVK for optimised eccentricity searches,
parameter estimation, and for injection campaigns to establish eccentricity-related observational selection effects.

Our findings further highlight that the exact distribution in eccentricity and frequency commonly resolvable by
LVK ($e_{10} \gtrsim 0.1$; \citep{Lower18}) should, in many cases, to leading order, follow a unified form regardless of the details of the underlying dynamical
formation channel. In other words, the binaries that enter the pinhole share the same final distribution regardless of their formation channel.
The obvious consequence is that, by definition, \textit{these different astrophysical formation mechanisms are fundamentally
indistinguishable in this range via their eccentricity alone}.
There are two options to rectify this problem. The first is to measure lower eccentricities, which may already be achievable for lower-mass
binaries and binaries with significant higher-mode content \citep{MooreYunes20, IRS25}, and will be routine with future
detectors \citep{Lower18, Saini24}.
At lower eccentricities, the distributions predicted for different dynamical or triple formation scenarios are
distinct \citep[e.g.,][]{Dorozsmai26, 2026arXiv260329019G}.
The second is to study eccentricity \textit{only} in correlation with other parameters.
Different formation scenarios might lead to different correlations
between eccentricity and redshift \citep{Dorozsmai26}, spin \citep{Samsing22, Fabj24, 2025arXiv251007952F, Stegmann25}, and mass. 

Finally, we acknowledge that our calculation assumes orbit-averaged $2.5 \ \rm{PN}$ level dynamics,
and is based solely on the equations in \cite{Peters:1964bc}.
This limit is known to be inaccurate in the extremely high eccentricity regime, in which the Keplerian
orbital elements are also ill-defined. In addition, we note that defining $e_f$ at the orbital frequency, $f_T$,
as recently proposed by \cite{2024ApJ...969..132V}, maps to highly relativistic encounters for even modest values of $e_f$,
which therefore requires full NR simulations to explore further.
The question of how and where to define eccentricity, therefore, still seems to be challenging and needs a revisit.
Regardless of these issues, the resulting eccentricity distribution should still follow a common form.

It is important to note that we do not aim to characterise the full population of eccentric mergers, but only those in the pinhole regime. Different dynamical formation channels contribute differently to the pinhole population, but all the binaries that enter the pinhole share the same population regardless of their formation channel, as shown in this \textit{Letter} using three independent methods: analytically, semi-analytically (continuity equation), and by comparing our results to numerical simulations of different dynamical channels. Entering the pinhole regime requires exceeding a critical eccentricity, and conditions on the geometry of the system, as discussed above.

\begin{acknowledgments}
MR gratefully acknowledges support from the Institute for Advanced Study, the Kovner Member Fund, and the Gruber Foundation Fellowship. I.M.R-S acknowledges support from the Science and Technology Facilities Council Ernest Rutherford Fellowship grant number UKRI2423.
JS is supported by the Villum Fonden grant No. 29466,
and by the ERC Starting Grant no. 101043143 –
BlackHoleMergs. The Center of Gravity is a Center of Excellence
funded by the Danish National Research Foundation under grant No. 184.
\end{acknowledgments}

\appendix

\section{Frequency dependence and general assembly distributions}\label{appendix:Jacobian}

In this appendix, we demonstrate another method to calculate the eccentricity distribution and its dependence on the frequency (orbital frequency in particular) using the continuity equation.  

Let $p_0(e_0)$ be the eccentricity distribution at assembly, in a set-up in which we assume that all the binaries are formed at a given small fixed orbital frequency $f_T = f_0$. We let the eccentricity propagate following the continuity equation, and measure it again after time $t$,

\begin{align}
\frac{\partial P(e,t)}{\partial t}+\frac{\partial}{\partial e}\left[\frac{de}{dt}P(e,t)\right]=0.
\end{align}

   The continuity equation leads us to the conclusion that segments in the probability distribution function $P(e)$ conserve their weight at different frequencies, and hence for any given orbital frequency $f$, which could be uniquely identified with a given corresponding time,  

\begin{align}
P(e)= P_0(e_0)\left|\frac{de_0}{de}\right|.
\end{align}

To calculate the Jacobian $|de_0/de|$, we define a function,  

\begin{align}
\Phi(e,e_0)=\frac{f}{f_0}=\left(\frac{a_0(e_0)}{a(e)}\right)^{3/2}
\end{align}

\noindent
and use the relation between $a$ and $e$ as given in eq. 5.11 \cite{Peters:1964bc}, and also discussed in Section \ref{sec:universal}. This function formulates an implicit relation between $e$ and $e_0$, and in the general case, an expression for $de_0/de$ could be extracted numerically.
This statement is general and holds for any initial eccentricity distribution, and allows us to calculate the new eccentricity distribution.

\begin{figure}[h]
    \centering
    \includegraphics[width=1\linewidth]
    {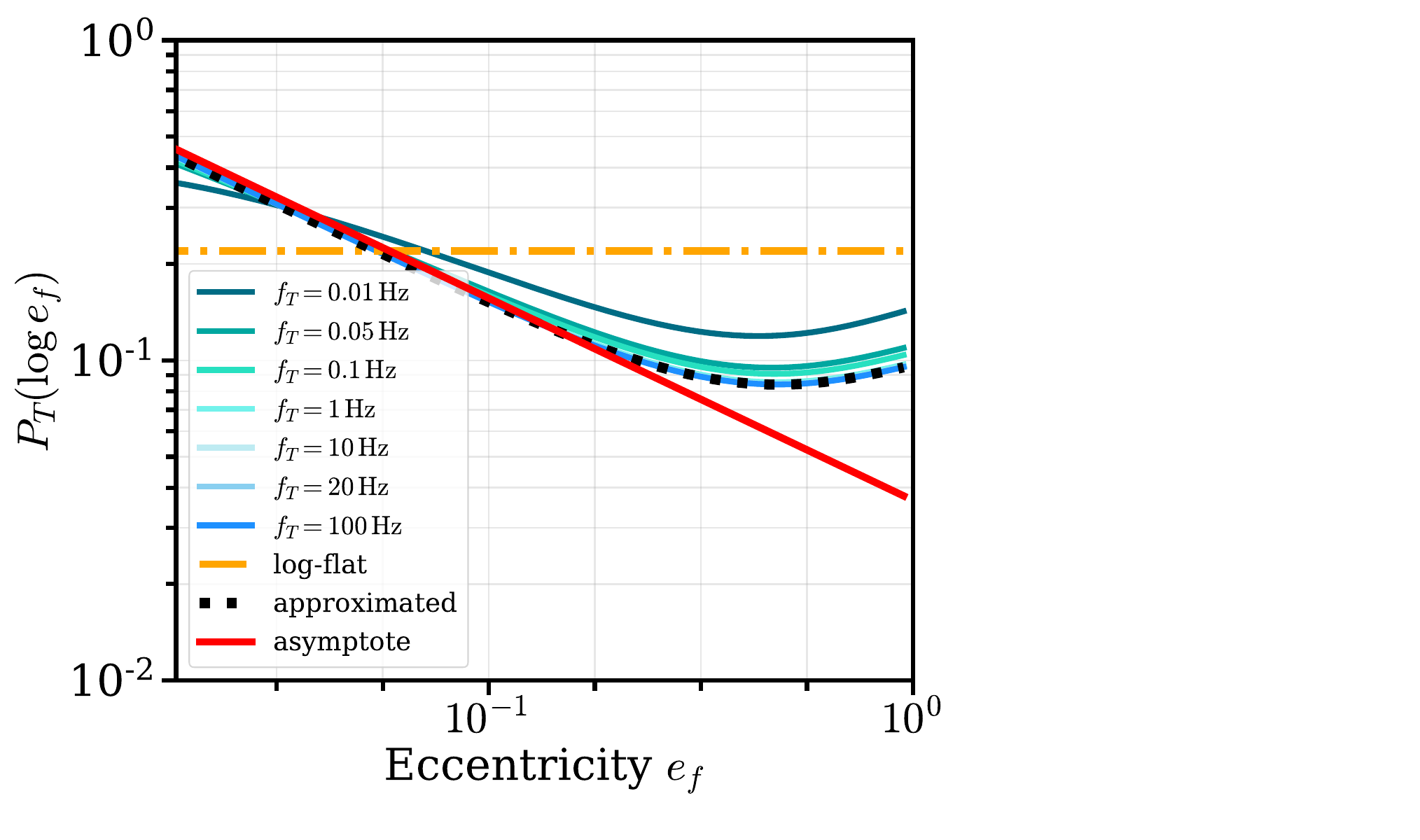}
    \caption{ The probability \textit{density} function of eccentricity, as calculated in Appendix \ref{appendix:Jacobian}, at different orbital frequencies (solid blue lines), compared to log-flat distribution (dotted-dashed line), approximated solution (Eq. \ref{eq:PT}, dotted) and asymptotic (solid red line). 
    }
    \label{fig:p(e)_analytical}
\end{figure}

Our results and comparisons between different probability \textit{density} functions are illustrated in Fig. \ref{fig:p(e)_analytical}. In solid blue lines, we illustrate the distribution function as calculated at different orbital frequencies, assembled with thermal distribution at an initial frequency $10^{-4} \ \rm{Hz}$, derived based on the calculation in Appendix \ref{appendix:Jacobian}, compared to the distribution described in Eq. \ref{eq:PT} (dotted), log-flat (dotted-dashed) and asymptotic (solid red). As shown, our closed-form solution in Eq. \ref{eq:PT} compares well with the results from the calculation in Appendix \ref{appendix:Jacobian}, and as the frequency increases, the latter converges to the former. The asymptotic solution describes the slope well, while the thermal and log-flat distributions differ substantially from our solution as described by the two methods above. It should be noted that in Fig. \ref{fig:p(e)_analytical}, we show a normalised probability density function as a function of the eccentricity $e_f$.

\section{Different definitions for frequency}\label{app:B}

\begin{figure}[h]
    \centering
    \includegraphics[width=.5 \linewidth]{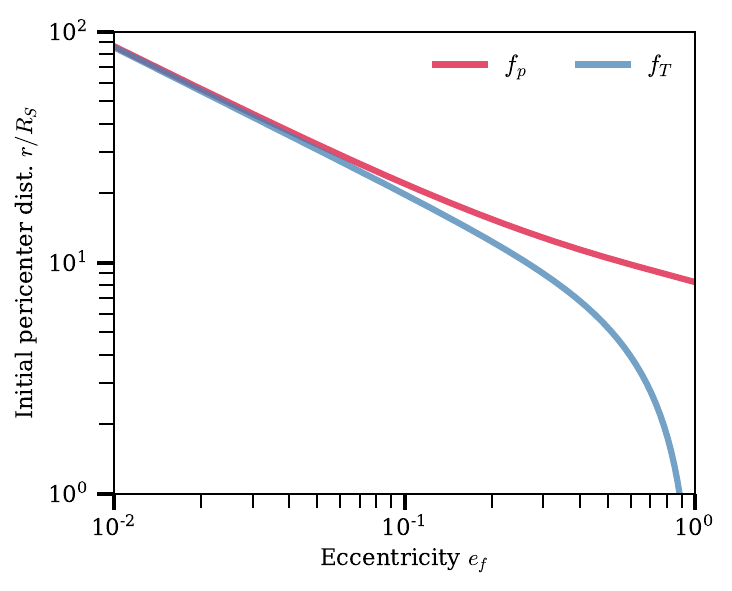}
    \caption{Initial pericenter distance scaled by $R_s = 2GM/c^2$, $r/R_s$, as a function of
    eccentricity, $e_f$, derived using Eqs. \eqref{eq:r_EM} and \eqref{eq:r_EM2}.
    Results for $f_p$ and $f_T$ are shown in {\it red} and {\it blue}, respectively. For this figure we
    have assumed $M=40M_{\odot}$ and $f=20~\text{Hz}$.}
    \label{fig:r_ecc}
\end{figure}

Fig. \ref{fig:r_ecc} shows the initial pericenter distance scaled by $R_s = 2GM/c^2$, $r/R_s$, as a function of eccentricity, $e_f$,
plotted using Eqs. \eqref{eq:r_EM} and \eqref{eq:r_EM2} for $f_p$ and $f_T$, respectively. For this we have assumed $M=40 \ M_{\odot}$ and $f=20\ \rm{Hz}$.
As seen, when using $f_T$ as reference frequency, $r/R_s$ drops steeply towards $0$ for increasing eccentricity, which calls for a full
NR treatment even for modest eccentricities. Finite-size effects, breakdown of the PN formalism, as well as ill-defined orbital elements,
make this an extremely challenging mapping that needs further work. We note that \cite{2024ApJ...969..132V}
recently proposed $f_T$ as the optimal choice for comparison, however, our work seems to suggest this problem has to be revisited.

\section{Observational Implications}\label{sec:Observational Implications}

Different dynamical channels and environments could be distinguished and identified below their critical eccentricity, as outlined
in Sec. \ref{sec:astro_examples}. Current detectors are sensitive to eccentricities $e_{\rm{10}}\gtrsim 0.1$ \citep{Lower18}, which could potentially lead to degeneracy between observed eccentricity distributions originating from the different dynamical channels discussed above. This degeneracy could be potentially resolved either by considering correlations of the eccentricity with other observables or by observing lower eccentricities, below the current sensitivity.

\begin{figure}
    \centering
    \includegraphics[width=\linewidth]{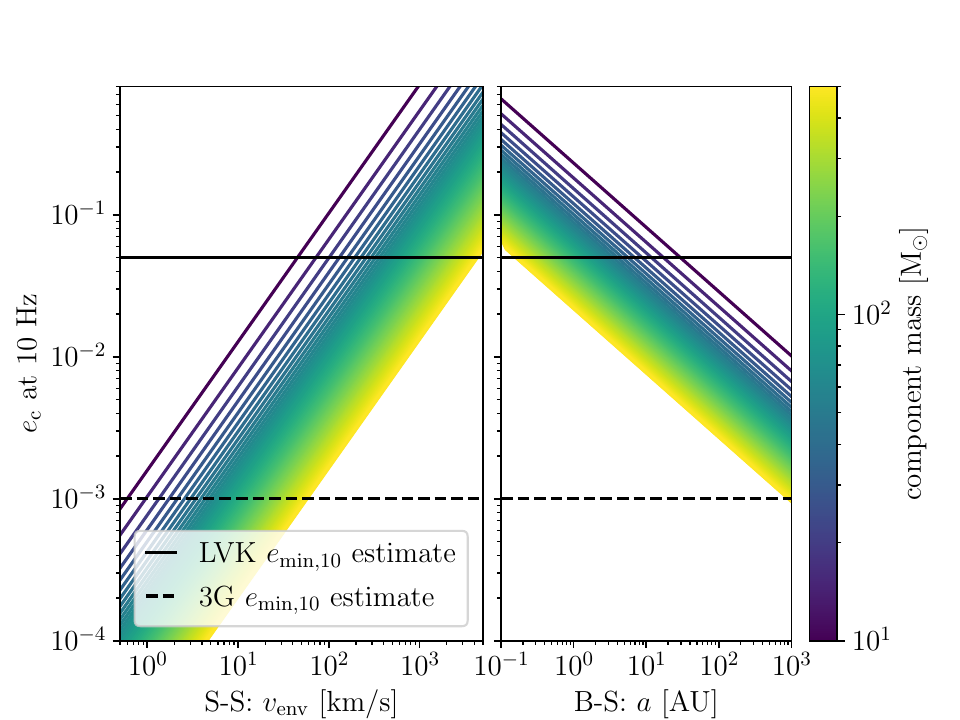}
    \caption{$e_\mathrm{crit}$ for varying $v_\mathrm{env}$ (left, single-single encounters; S-S) and binary semi-major axis $a$ (right, binary-single encounters; B-S) coloured by component mass, calculated using Eqs. \eqref{eq:ec_venv} and \eqref{eq:ec_BS}. Estimated minimum measurable eccentricities for GW150914-like systems in current LVK detectors and future 3G detectors are shown with solid and dashed horizontal lines, respectively \citep{Lower18,Saini24}. (We note that eccentricity measurability is a function of SNR \citep{Singh2025}, so these horizontal lines are purely indicative.)  With 3G detectors, eccentricity measurability is sufficient that $e_\mathrm{crit}$ is measurable for a wide range of environment velocities, binary separations, and component masses.}
    \label{fig:observational_implications}
\end{figure}

In Fig. \ref{fig:observational_implications} we introduce the critical eccentricity to enter the pinhole regime at $10 \ \rm{Hz}$, $e_{\rm{crit}}$, for different 
masses (different colours), as a function of the characteristic velocity of the environment, $v_{\rm{env}}$, for merger channels that are driven by single-single interactions (left panel), and the separation, for merger channels driven by binary-single encounters  (right panel).

\bibliography{sample701}{}
\bibliographystyle{aasjournalv7}




\end{document}